\documentclass[twocolumn,aps,pre]{revtex4}

\usepackage[koi8-r]{inputenc}
\usepackage[english,russian]{babel}
\usepackage{amsmath,amssymb,graphicx}
\usepackage{xcolor}
\definecolor{yblue}{rgb}{0.06, 0.3, 0.57}

\begin{document}
\title{Численное моделирование световых структур в объемных ENZ-средах
с керровской нелинейностью}

\author{В.\,П.\,Рубан}
\email{ruban@itp.ac.ru}
\affiliation{Институт теоретической физики им.~Л.\,Д.\,Ландау РАН, 142432, 
Черноголовка, Россия}

\date{\today}

\begin{abstract}
Предложена упрощенная математическая модель для описания динамики 
квазимонохроматической световой волны в объеме эффективно изотропного
метаматериала с близкой к нулю усредненной диэлектрической проницаемостью
(т. н. ENZ-среда) при наличии слабой пространственной неоднородности,
керровской нелинейности, а также линейного усиления под действием 
внешней накачки. Модель представляет собой векторное уравнение 
Гинзбурга-Ландау общего вида с доминированием слагаемого ротор-ротор
в дисперсионном операторе и напоминает уравнение для электромагнитных
волн в плазме [Е. А. Кузнецов, 1974]. В случае чисто действительных
керровских коэффициентов хорошо работает численный метод с расщепленным
шагом (split-step Fourier method), который позволил пронаблюдать различные
варианты нетривиальной эволюции как центрально симметричных, так и
тороидальных векторных волновых структур в квадратичной потенциальной
яме, а также нелинейное взаимодействие между продольными и поперечными
волнами в случае комбинации указанных структур.
\vspace{1mm}

\end{abstract}

\maketitle
%%%%

{\bf Введение.}
За последние полтора-два десятилетия в оптических исследованиях огромное
внимание стало уделяться так называемым ENZ-средам (Epsilon Near Zero) ---
искусственно созданным из элементов суб-волнового размера $a_s$ композитным
материалам, в которых некоторые собственные числа вещественной части тензора
эффективной диэлектрической проницаемости $\varepsilon_{jl}(\omega)$ принимают
очень малые значения в области частот видимого либо инфра-красного спектра
\cite{realization2013,IZdielectric2013,LE2017,materials2019,photonics2021}. 
Более того, в типичной ситуации соответствующая функция $\varepsilon'(\omega)$
проходит через ноль при некоторой частоте $\omega_0$. Распространение света
в подобной среде обладает целым рядом необычных свойств, включая возможность
сильно нелинейных режимов
\cite{compos-1,compos-2,compos-3,compos-4,nonlin_effects2019,nonlin_optics2024}.
Именно о нелинейной динамике оптического поля и пойдет далее речь.

Надо сказать, что большинство применений ENZ-метаматериалов ориентировано на
тонкие слои/пленки, в которых тензор $\varepsilon_{jl}(\omega)$ анизотропен,
но имеется также интерес к поведению света в объемных ENZ-образцах (см., например,
\cite{compos-3,compos-4,pulse2016,frozen2016}). В этой связи возникает необходимость
иметь удобную  математическую модель для адекватного описания оптических волн 
по крайней мере в ENZ-средах простейшего (с теоретической точки зрения) типа 
--- эффективно изотропных. До сих пор такая модель не была предложена. 
Цель данной работы  --- частично заполнить этот пробел в теории, рассмотрев
квазимонохроматическую волну в идеализированной изотропной ENZ-среде,
обладающей только кубической нелинейностью керровского вида. 

{\bf Феноменологическая модель.}
Если ${\bf E}({\bf r},t)$  --- медленная комплексная временная огибающая 
(усредненного по суб-волновым неоднородностям) электрического поля,
то соответствующий нелинейный вклад в правую часть волнового уравнения
\begin{equation}
c^2\mbox{rot}\,\mbox{rot}{\bf E}= (\omega_0+i\partial_t)^2{\bf D}
\label{main_wave_eq}
\end{equation}
мы примем равным 
\begin{equation}
(\omega_0+i\partial_t)^2{\bf D}^{(3)}\approx\omega_0^2[\alpha|{\bf E}|^2{\bf E}
+\beta({\bf E}\cdot{\bf E}){\bf E}^*],
\end{equation}
где $\alpha=\alpha'+i \alpha''$ и $\beta=\beta'+i\beta''$ --- два керровских
коэффициента. Здесь подразумевается, что интенсивность света достаточно мала,
так что $|\alpha|E^2\ll 1$, где $E\equiv|{\bf E}|$. Линейный по электрическому
полю вклад в правую часть уравнения (\ref{main_wave_eq}) мы получим разложением
эффективного тензора диэлектрической проницаемости $\varepsilon_{jl}(\omega,{\bf k})$
по малому отклонению частоты $(\omega-\omega_0)$ и малому волновому вектору
${\bf k}$ (феноменологически предполагая, что такое разложение имеет место). 
Мы сейчас рассматриваем среду без искусственной хиральности, поэтому
в разложении не будет линейных по ${\bf k}$ членов. Таким образом, с учетом
изотропности и возможной слабой пространственной неоднородности среды имеем формулу
\begin{eqnarray}
\varepsilon_{jl}&\approx& \delta_{jl} [\tau(\omega-\omega_0) +i\gamma-u({\bf r})]
\nonumber\\
&-&(D_\perp-i\nu_\perp)[k^2\delta_{jl}- k_j k_l] -(D_\parallel-i\nu_\parallel)k_j k_l,
\end{eqnarray}
где $\gamma$ -- параметр поглощения, $u({\bf r})$ --- неоднородность, а также введены
феноменологические вещественные константы $\tau\sim 1/\omega_0$, $D_\perp\sim a_s^2$,
$D_\parallel\sim a_s^2$, $\nu_\perp\sim a_s^2$ и $\nu_\parallel\sim a_s^2$.
Оценки для этих дисперсионных и ``вязких'' параметров приведены по соображениям
размерности, что не исключает сильно отличающихся от единицы индивидуальных
численных коэффициентов (по поводу эффектов пространственной дисперсии в
метаматериалах см. \cite{nonlocal2007,nonlocal2009,nonlocal2015,nonlocal2025}). 
Более того, как показывает теория бесстолкновительной плазмы, в некоторых средах
диссипативные параметры $\nu_\perp$ и $\nu_\parallel$ в действительности могут
оказаться неаналитическими функциями от $k^2$ с нулевой асимптотикой при $k\to 0$
\cite{Theor_Phys_10}. Но для твердых материалов с металлическими включениями такой
вариант выглядит маловероятным.

Диссипативные константы $\gamma$, $\nu_\perp$ и $\nu_\parallel$ приводят
к затуханию волны. Чтобы противодействовать этому процессу, можно добавить
в среду активные центры и накачивать их внешним источником. В результате
получается скомпенсированная по потерям либо даже слегка неустойчивая среда
с эффективным коэффициентом линейного усиления $G$ вместо поглощения $\gamma$
(см., например, \cite{RDFC2011,active_2010,active_2011,SYG2013}).

С учетом вышесказанного, для линейного вклада в правую часть волнового уравнения
(\ref{main_wave_eq}) получается несколько непривычная формула, в которой все
слагаемые --- одного порядка малости:
\begin{eqnarray}
&&(\omega_0+i\partial_t)^2{\bf D}^{(1)}\approx
\omega_0^2[i(\tau {\bf E}_t-G{\bf E}) -u({\bf r}){\bf E}
\nonumber\\
&&\quad -(D_\perp-i\nu_\perp)\,\mbox{rot}\,\mbox{rot}{\bf E}
+(D_\parallel-i\nu_\parallel)\nabla(\nabla\cdot{\bf E})].
\end{eqnarray}
Далее суммируем линейный и нелинейный вклады, попутно выбирая новый масштаб
времени $Q^2\tau$, новый масштаб длины $Qc/\omega_0$ и новый масштаб 
электрического поля $1/(Q\sqrt{\alpha'})$, где $Q$ --- некоторое
большое число (практически удобно подразумевать $Q\sim 50$). 
С учетом неравенства $(c/\omega_0)\gg a_s$, которое позволяет пренебречь
$D_\perp$, мы приходим в результате к векторному уравнению Гинзбурга-Ландау
\begin{eqnarray}
i({\bf E}_t- G{\bf E})=(1-i \nu_\perp)\,\mbox{rot}\,\mbox{rot}{\bf E}
-(D_\parallel-i \nu_\parallel)\nabla(\nabla\cdot{\bf E})&&\nonumber\\
+U({\bf r}){\bf E}
-(1+i\sigma)|{\bf E}|^2{\bf E}
-\eta({\bf E}\cdot{\bf E}){\bf E}^*,\qquad&&
\label{GL_eq}
\end{eqnarray}
где все параметры и само поле ${\bf E}$ теперь безразмерные, причем
$\eta=\beta/\alpha'$ в композитных материалах может принимать
значения в широком диапазоне \cite{compos-1,compos-2,compos-3,compos-4}.
Продольный дисперсионный параметр $D_\parallel$ обязан быть по крайней мере
на два-три порядка меньше единицы. То же самое предположительно верно и для
диффузионных/вязких коэффициентов $\nu_\perp$, $\nu_\parallel$.

Как и следовало ожидать, наша модель является обобщением известного уравнения
для электромагнитных волн в плазме \cite{K1974}, поскольку сама по себе плазма
является примером нелинейной ENZ-среды (см. также 
\cite{ZMR1985,KRZ1986,PSSW1991,HS2009} и ссылки там).
Мы видим, что продольные оптические волны (аналог ленгмюровских волн в плазме)
с необходимостью должны быть включены в рассмотрение в ENZ-среде. Продольное 
оптическое поле при этом оказывается во много раз ``массивнее'' поперечного
в меру малости параметра $D_\parallel$.

Наличие в векторном уравнении (\ref{GL_eq}) нескольких параметров, а также свобода
в выборе ``потенциала'' $U({\bf r})$ заставляют предполагать богатую на различные
режимы нелинейную динамику, включая волновые коллапсы. 

{\bf Численный метод.}
Весьма удачно, что уравнение (\ref{GL_eq}) разрешено относительно временной
производной. Это позволяет производить компьютерные расчеты эволюции системы
хорошо проверенными численными методами. Мы в данной работе ограничимся
рассмотрением захваченных волн в изотропной квадратичной яме $U=w (x^2+y^2+z^2)$.
Кроме того, для удобства численного моделирования будут предполагаться чисто
действительные керровские коэффициенты, то есть $\sigma=0$, 
$\mbox{Im}\, \eta =0$. В таком случае можно применить метод Фурье с
расщепленным шагом по времени второго порядка аппроксимации, который по
вычислительным затратам сравним в методом первого порядка. Важным моментом
здесь является наличие явного аналитического решения для локальной системы
обыкновенных дифференциальных уравнений
\begin{equation}
id{\bf E}/dt=U{\bf E} -|{\bf E}|^2{\bf E}-\eta({\bf E}\cdot{\bf E}){\bf E}^*.
\label{dynam_local}
\end{equation}
Решение задачи Коши находится легко и определяется следующей формулой,
в которой использованы обозначения $I_0=|{\bf E}_0|^2$,
$S_0=({\bf E}_0\cdot{\bf E}_0)$, $\lambda=\eta\sqrt{I_0^2-|S_0|^2}$ и 
${\bf R}_0=i\eta(S_0{\bf E}^*_0 -I_0{\bf E}_0)$:
\begin{equation}
{\bf E}(t)=\Big[{\bf E}_0\cos(\lambda t)
+{\bf R}_0\frac{\sin(\lambda t)}{\lambda}\Big]e^{i(1+\eta)I_0 t-iUt}.
\label{local}
\end{equation}
Очевидно, что в системе (\ref{GL_eq}) дополняющей динамикой поля ${\bf E}$
по отношению к локальному вкладу (\ref{dynam_local}) является дисперсионная
эволюция
\begin{equation}
i{\bf E}_t=i G {\bf E}+(1-i \nu_\perp)\,\mbox{rot}\,\mbox{rot}{\bf E}
-(D_\parallel-i\nu_\parallel)\nabla(\nabla\cdot{\bf E}),
\end{equation}
для которой решение задачи Коши имеет простой вид в Фурье-представлении:
\begin{eqnarray}
{\bf E}_{\bf k}(t)=
\frac{{\bf k}({\bf k}\cdot{\bf E}_{\bf k}(0))}{k^2}
\exp\{[G-(iD_\parallel+\nu_\parallel)k^2] t\}\qquad && \nonumber\\
+\Big[{\bf E}_{\bf k}(0)-\frac{{\bf k}({\bf k}\cdot{\bf E}_{\bf k}(0))}{k^2}\Big]
\exp\{[G-(i+\nu_\perp)k^2] t\}.&&
\label{disp}
\end{eqnarray}
Для выполнения этого нелокального дисперсионного преобразования поля 
${\bf E}({\bf r})$ необходимо проделать сначала прямое, а затем обратное
дискретное преобразование Фурье для каждой компоненты вектора ${\bf E}$.
Формула (\ref{disp}) легко обобщается на случай, когда $\nu_\perp$ и 
$\nu_\parallel$ являются функциями от $k$, а также когда имеется два различных
значения коэффициента усиления $G_\parallel$ и $G_\perp$ соответственно 
для продольных и поперечных волн.

В методе Фурье с расщепленным шагом второго порядка точности продвижение
по времени на $M\gg 1$ шагов $\Delta t$ производится в следующем порядке:

1. Выполняется локальное преобразование (\ref{local}) на пол-шага $0.5(\Delta t)$,
со своими ${\bf E}_0$ и $U$ на каждом узле решетки.

2. Выполняется последовательность из $(M-1)$ пары преобразований 
дисперсионное--локальное с полным шагом каждое.

3. Выполняется дисперсионное преобразование с полным шагом.

4. Выполняется локальное преобразование на пол-шага.

В результате система эволюционирует во времени на интервал $M (\Delta t)$
и при этом достигается общая точность порядка $(\Delta t)^2$. 
Здесь уместно упомянуть для полноты картины, что случай
насыщающейся нелинейности, когда $\alpha$ и $\beta$ являются функциями от
вещественных скалярных аргументов $|{\bf E}|^2$ и $|({\bf E}\cdot{\bf E})|^2$,
рассматривается аналогично.

В представленных далее нескольких численных экспериментах параметр $\eta=0.5$,
что типично для обычных диэлектриков с быстрым нелинейным откликом.
Если керровские коэффициенты имеют мнимую часть, то можно применять
псевдо-спектральный метод с временным шагом по схеме Рунге-Кутта 4-го порядка.
При том же значении $\Delta t$ вычисления потребуют примерно в четыре
раза больше времени.

Вычислительная область представляла собой куб размера $2\pi$ с периодическими
граничными условиями. Дискретизация осуществлялась решеткой с 
$N=192\times 192\times 192$ узлами. При выполнении дисперсионного
преобразования обнулялись все Фурье-гармоники с $k>50$. С таким разрешением
типичный численный эксперимент занимал несколько суток машинного времени
на современном персональном компьютере, рассчитывая при этом эволюцию
системы до времен $t\sim 300$.

Начало координат помещалось в центре куба. Поскольку функция $w r^2$ при
периодическом продолжении претерпевает нежелательный излом на границе куба,
вместо нее использовался более сглаженный на краях вариант потенциальной ямы
$$
wr^2 \approx w[9-0.2\ln(1+\exp[-5(r^2-9)])].
$$

Наиболее тонкий вопрос --- выбор диссипативных параметров $\nu_\perp$ и
$\nu_\parallel$. Не имея под рукой экспериментальных данных о каком-либо
эффективно изотропном ENZ-метаматериале, мы далее будем использовать
произвольные значения в диапазоне от 0.001 до 0.01. Очевидно при этом, 
что для численного моделирования на долгие времена слишком малая диссипация
не очень привлекательна. Но надо сказать, что для тестирования численного кода
брались нулевые значения $\nu_\perp=0$ и $\nu_\parallel=0$, когда (при $G=0$)
сохраняется волновое действие ${\cal A}=\int|{\bf E}|^2 d{\bf r}$, а также
соответствующий гамильтониан системы
\begin{eqnarray}
&&{\cal H}=\int \Big[|\mbox{rot}\,{\bf E}|^2
+D_\parallel |(\nabla\cdot{\bf E})|^2 +U({\bf r})|{\bf E}|^2\nonumber\\
&&\qquad\qquad-\frac{1}{2}\big(|{\bf E}|^4+\eta|({\bf E}\cdot{\bf E})|^2\big)
\Big] d{\bf r}.
\end{eqnarray}

\begin{figure}%[htb]
\begin{center}
\includegraphics[width=0.95\columnwidth]{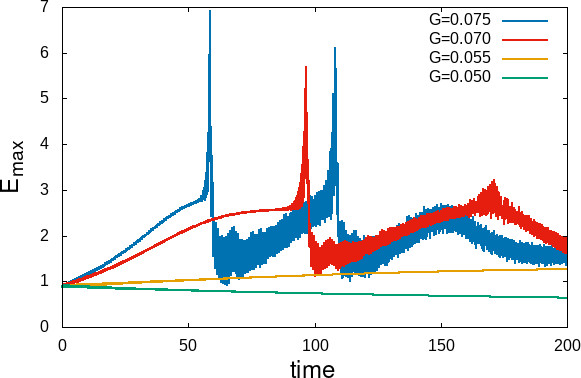}
\caption{
Эволюция максимального значения амплитуды электрического поля
для тороидальных структур при различных значениях поперечного
коэффициента усиления.
}
\label{E_max_bubliki}
\end{center}
\end{figure}

{\bf Примеры эволюции нелинейных структур.}
Выбирая параметры и начальные условия для демонстрационных примеров, прежде всего
обратим внимание, что при наличии потенциала $U=wr^2$ в бездиссипативном пределе
имеется два типа интересных мало-амплитудных решений. Первый тип --- это
потенциальные, центрально симметричные поля вида
\begin{equation}
{\bf E}_\parallel \propto {\bf r}\exp\Big[-\sqrt{(w/D_\parallel)}r^2/2\Big].
\label{pot_solution}
\end{equation}
Для краткости подобные конфигурации принято называть ``ежами''. 
Второй тип линейных решений --- соленоидальные, азимутально направленные 
поля в форме ``бублика'',
\begin{equation}
{\bf E}_\perp\propto [{\bf n}_z\times{\bf r}]\exp[-\sqrt{w}r^2/2].
\label{solenoid_solution}
\end{equation}
Отметим, что размеры локализации центрально симметричных и тороидальных решений
отличаются в ${D_\parallel}^{1/4}$ раз. Например, при $D_\parallel=0.01$ ``бублик''
оказывается в $\sqrt{10}\approx 3$ раза больше.

Интересно взять эти решения в качестве отправной точки и применить усиление $G$.
При медленном накоплении волнового действия ${\cal A}$ происходит постепенная 
деформация структуры за счет включения нелинейности. Размер сгустка оптического
поля при этом слегка уменьшается, что повышает темп ``вязкой'' диссипации. 
Достаточно сильное увеличение ${\cal A}$ должно привести к потере устойчивости. 
Какие моды окажутся наиболее неустойчивыми и в какой новый динамический режим
перейдет система --- зависит от ее параметров.

\begin{figure}%[htb]
\begin{center}
\includegraphics[width=0.46\columnwidth]{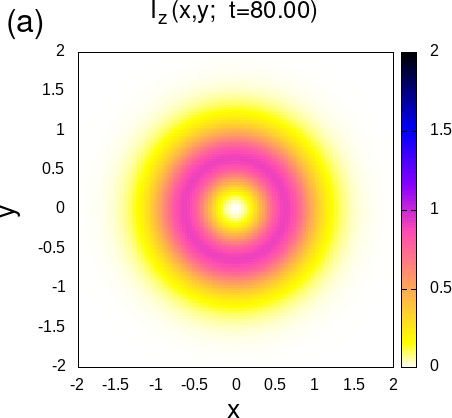}
\phantom{qq}
\includegraphics[width=0.46\columnwidth]{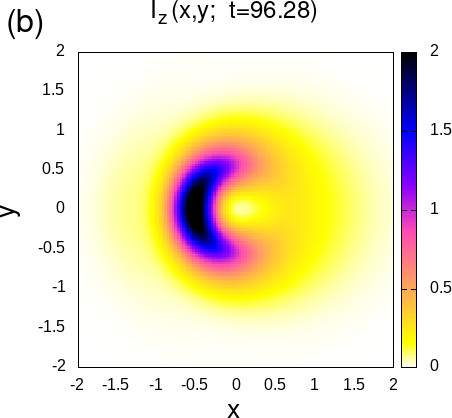}
\caption{
Примеры конфигураций ``бублика'': a) до потери устойчивости; b) в момент
сильной дипольной деформации во время всплеска колебаний. Показана
усредненная по $z$ интенсивность поля $I_z(x,y)=(1/2\pi)\int E^2 dz$
в два различных момента времени.
}
\label{dipole_xy}
\end{center}
\end{figure}

\begin{figure}%[htb]
\begin{center}
\includegraphics[width=0.46\columnwidth]{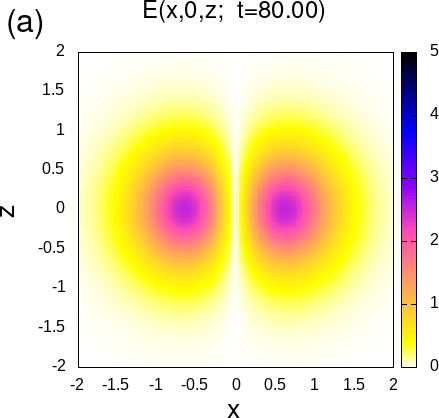}
\phantom{qq}
\includegraphics[width=0.46\columnwidth]{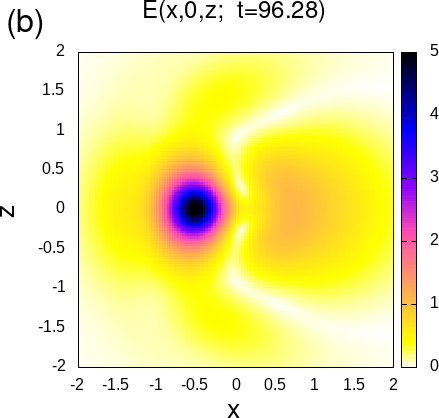}
\caption{
Примеры конфигураций ``бублика'' до потери устойчивости и во время всплеска:
амплитуда поля в плоскости $y=0$.
}
\label{dipole_xz}
\end{center}
\end{figure}

В первой серии численных экспериментов моделировалась эволюция слабо-возмущенных 
тороидальных структур. На рисунке \ref{E_max_bubliki} представлены зависимости
максимальной амплитулы поля $E_{\rm max}(t)$ для нескольких значений поперечного
параметра усиления $G_\perp$ при выборе остальных параметров $w=4.0$, 
$D_\parallel =0.01$, $\nu_\perp=\nu_\parallel =0.01$. При небольшом усилении
$G_\perp\lesssim 0.6$ происходил умеренный рост с насыщением либо даже спадание
амплитуды ``бублика''. В этих случаях диссипация ограничивала экспоненциальный
рост волнового действия и тем самым поддерживала баланс стабильной
слабонелинейной структуры, а то и вовсе оставалась преобладающей.
Но при большем усилении ``бублик'' оставался устойчивым лишь до определенного
момента, а затем на фоне тороидальной конфигурации развивалась неустойчивость,
что приводило к резкому всплеску колебаний. Обычно возбуждалась дипольная мода,
когда ``бублик'' попеременно утолщался то с одной стороны, то с противоположной. 
В ходе развития неустойчивости достигались достаточно мелкие масштабы,
на которых волновое действие быстро ``съедалось'' диссипацией, так что
система ``падала'' в докритическое состояние, хотя и с возбуждением различных
мод (с преобладанием дипольной моды). 
Примеры конфигураций до потери устойчивости и в момент всплеска показаны на 
рисунках \ref{dipole_xy}-\ref{dipole_xz} для случая $G_\perp = 0.70$.
В дальнейшем процесс накопления действия и всплеск дипольной моды могли повториться,
как в случае $G_\perp = 0.75$. Другой возможный сценарий --- после повторного
накопления действия не случилось второго сильного всплеска, но доминирующее
возбуждение перешло от дипольной моды к квадрупольной, как это имело место при 
$G_\perp = 0.70$ на временном интервале примерно между $t=170$ и $t=180$.

\begin{figure}%[htb]
\begin{center}
\includegraphics[width=0.95\columnwidth]{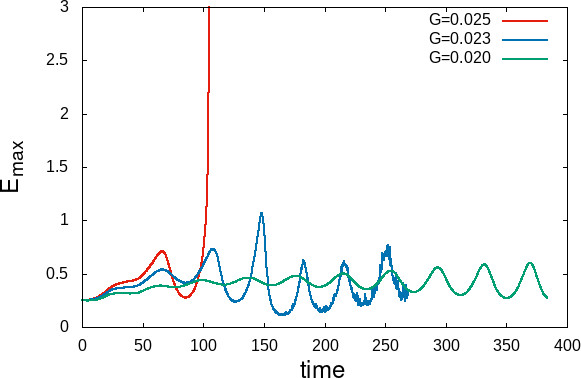}
\caption{
Максимальное значение амплитуды электрического поля в зависимости от времени
для центрально симметричных структур при различных значениях накачки.
}
\label{E_max_ezhiki}
\end{center}
\end{figure}

Во второй серии численных экспериментов моделировалась динамика ``ежей''.
Были выбраны параметры $w=0.25$, $D_\parallel =0.01$, 
$\nu_\perp=\nu_\parallel =0.001$. Максимальная амплитуда поля
$E_{\rm max}(t)$ для нескольких значений коэффициента усиления 
представлена на рисунке \ref{E_max_ezhiki}. В отличие от случая ``бубликов'',
умеренно нелинейные долгоживущие ``ежи'' существуют в колебательном режиме.
При этом амплитуда их радиальных пульсаций сначала постепенно нарастает,
затем относительно быстро спадает, а далее снова начинается рост.
Начиная с некоторого значения коэффициента усиления, в системе происходит
настолько резкий рост амплитуды и настолько сильное пространственное
сжатие сгустка оптического поля, что можно говорить о тенденции к коллапсу.
Разумеется, нашим численным методом невозможно в точности ответить на вопрос
о возможности коллапса в диссипативной системе (\ref{GL_eq}).

\begin{figure}%[htb]
\begin{center}
\includegraphics[width=0.46\columnwidth]{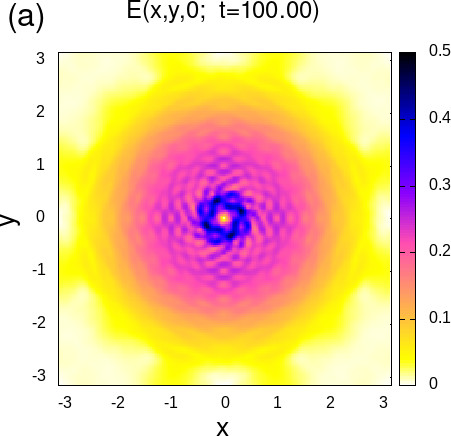}
\phantom{qq}
\includegraphics[width=0.46\columnwidth]{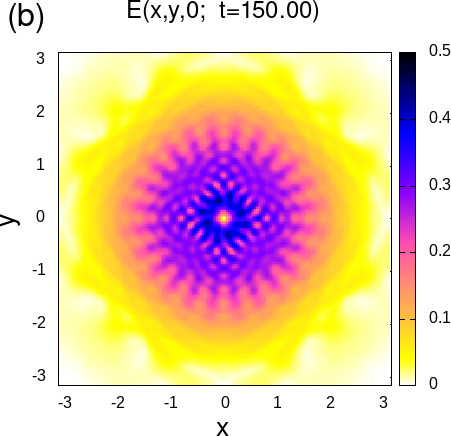}\\
\includegraphics[width=0.46\columnwidth]{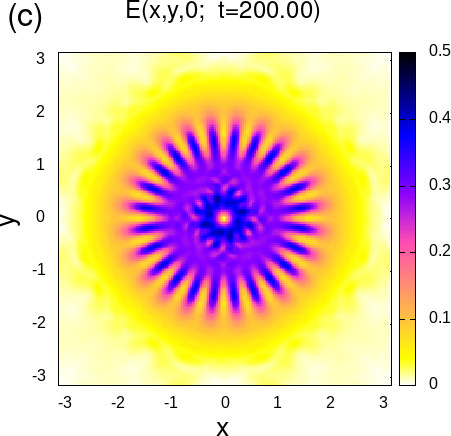}
\phantom{qq}
\includegraphics[width=0.46\columnwidth]{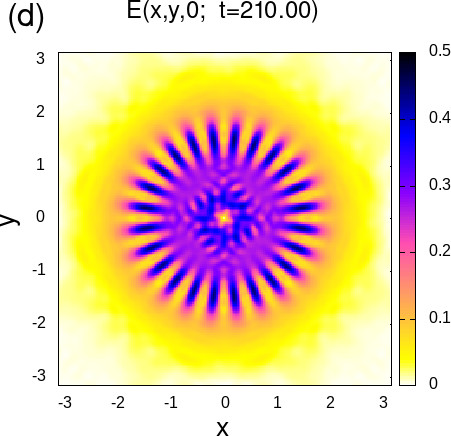}
\caption{
Пример развития волновых узоров при нелинейном взаимодействии продольных и 
поперечных волн. 
}
\label{textures}
\end{center}
\end{figure}

В третьей серии численных примеров начальное состояние было суперпозицией 
``ежа'' и ``бублика''. Весьма примечательно, что нелинейное взаимодействие
сравнимых по амплитуде продольных и поперечных полей приводило в этих случаях
к образованию мелкомасштабных, но все еще достаточно неплохо разрешаемых
числовой решеткой волновых ``узоров'', вроде тех, что показаны на рисунке
\ref{textures}. В данном примере $w=1.0$, $D_\parallel =0.01$, а эффекты
диссипации и накачки отсутствуют, то есть $\nu_\perp=\nu_\parallel=G=0$. 
Волновое действие ${\cal A}$ и гамильтониан ${\cal H}$ здесь сохранялись 
с точностью до пяти десятичных знаков, что свидетельствует о хорошем
качестве применяемого численного метода.
Теоретическая интерпретация подобных текстур пока отсутствует.

{\bf Заключение.} Таким образом, в данной работе приведены доводы в пользу
применимости векторного уравнения Гинзбурга-Ландау для описания эволюции
световой волны в объеме эффективно изотропного ENZ-метаматериала с керровской
нелинейностью. Реализован надежный численный метод, который на нескольких
примерах продемонстрировал весьма интересную динамику локализованных световых
структур. Локализация света в наших численных экспериментах достигалась за
счет слабой пространственной неоднородности среды. Надо сказать, что в
однородном случае все попытки автора численно осуществить самоподдерживающиеся
векторные структуры пока не увенчались успехом. Более того, теория волнового
коллапса в консервативных системах с квадратичной дисперсией и кубической
нелинейностью (как раз наш случай) говорит о неустойчивости любых
самолокализованных трехмерных равновесных конфигураций, по крайней мере по
отношению к изменению масштаба \cite{K1974,ZMR1985,KRZ1986}. 
Эти соображения приводят нас к сомнению относительно устойчивости найденного
в работе \cite{frozen2016} ``бубличного'' решения некоей модели, 
которая заведомо попадает в область применимости приближенного уравнения
(\ref{GL_eq}). Но при наличии усиления и вязкой диссипации вопрос о
возможности самоорганизации света в однородной ENZ-среде остается открытым.

На рисунках в этой краткой работе представлена далеко не вся полезная
информация, которую можно извлечь из численного моделирования.
Например, всплеск колебаний дипольной моды при достижении кризисного
значения волнового действия сопровождается генерацией потенциального поля.
В будущем было бы полезно более детально визуализировать векторную
структуру этого нестационарного, пространственно трехмерного решения.

В целом, дальнейшее исследование динамической системы (\ref{GL_eq})
при различных значениях параметров выглядит весьма перспективным делом. 

{\bf Финансирование работы.}
Работа выполнена в рамках госзадания по теме FFWR-2024-0013.

{\bf Конфликт интересов.}
Автор данной работы заявляет, что у него нет конфликта интересов.


\begin{thebibliography}{99}

\bibitem{realization2013} R. Maas, J. Parsons, N. Engheta, and A. Polman,
%Experimental realization of an epsilon-near-zero metamaterial at visible wavelengths,
Nature Photonics {\bf 7}, 907 (2013).

\bibitem{IZdielectric2013} P. Moitra, Y. Yang, Z. Anderson, I. I. Kravchenko,
D. P. Briggs, and J. Valentine,
% Realization of an all-dielectric zero-index optical metamaterial,
Nature Photonics {\bf 7}, 791 (2013).

\bibitem{LE2017} I. Liberal and N. Engheta,
%Near-zero refractive index photonics
Nature Photonics {\bf 11}, 149 (2017).

\bibitem{materials2019}
N. Kinsey, C. DeVault, A. Boltasseva, and V. M. Shalaev, 
%Near-zero-index materials for photonics, 
Nature Reviews Materials {\bf 4}, 742 (2019).

\bibitem{photonics2021}
J. Wu, Z. T. Xie, Y. Sha, H. Y. Fu, and Q. Li,
%Epsilon-near-zero photonics: infinite potentials,
Photonics Research {\bf 9}(8), 1616 (2021).

\bibitem{compos-1} J. E. Sipe and R. W. Boyd, 
%Nonlinear susceptibility of composite optical materials in the Maxwell Garnett model,
Phys. Rev. A {\bf 46}, 1614 (1992).

\bibitem{compos-2}
G. L. Fischer,  R. W. Boyd, R. J. Gehr, S. A. Jenekhe,
J. A. Osaheni, J. E. Sipe, and L. A. Weller-Brophy,
%Enhanced Nonlinear Optical Response of Composite Materials,
Phys. Rev. Lett. {\bf 74}, 1871 (1995).

\bibitem{compos-3} A. Ciattoni, C. Rizza, and E. Palange,
%Extreme nonlinear electrodynamics in metamaterials with very small
%linear dielectric permittivity,
Phys. Rev. A {\bf 81}, 043839 (2010).

\bibitem{compos-4} C. Rizza, A. Ciattoni, and E. Palange,
%Two-peaked and flat-top perfect bright solitons in nonlinear metamaterials
%with epsilon near zero,
Phys. Rev. A {\bf 83}, 053805 (2011).

\bibitem{nonlin_effects2019}
O. Reshef, I. De Leon, M. Z. Alam, and R. W. Boyd,
%Nonlinear optical effects in epsilon-near-zero media
Nature Reviews Materials {\bf 4}, 535 (2019).

\bibitem{nonlin_optics2024}
D. Fomra, A. Ball, S. Saha, J. Wu, Md. Sojib, A. Agrawal, H. J. Lezec, and N. Kinsey,
%Nonlinear optics at epsilon near zero: From origins to new materials,
Applied Physics Reviews {\bf 11}(1), 011317 (2024).

\bibitem{pulse2016}
A. Ciattoni, C. Rizza, A. Marini, A. Di Falco, D. Faccio, and M. Scalora,
%Enhanced nonlinear effects in pulse propagation through epsilon-near-zero media,
Laser Photonics Rev. {\bf 10}(3),  517 (2016).

\bibitem{frozen2016}
A. Marini and F. J. Garcia de Abajo,
%Self-organization of frozen light in near-zero-index media with cubic nonlinearity,
Scientific Reports {\bf 6}, 20088 (2016).

\bibitem{nonlocal2007}
J. Elser, V. A. Podolskiy, I. Salakhutdinov, and I. Avrutsky,
%Nonlocal effects in effective-medium response of nanolayered metamaterials,
Appl. Phys. Lett. {\bf 90}, 191109 (2007).

\bibitem{nonlocal2009} 
R. J. Pollard, A. Murphy, W. R. Hendren, P. R. Evans, R. Atkinson,
G. A. Wurtz, A. V. Zayats, and V. A. Podolskiy, 
%Optical nonlocalities and additional waves in epsilon-near-zero metamaterials,
Phys. Rev. Lett. {\bf 102}, 127405 (2009).

\bibitem{nonlocal2015} A. Ciattoni and C. Rizza,
% Nonlocal homogenization theory in metamaterials:
%Effective electromagnetic spatial dispersion and artificial chirality,
Phys. Rev. B {\bf 91}, 184207 (2015).

\bibitem{nonlocal2025} 
F. Monticone {\it et al.} (67 authors), 
%Nonlocality in photonic materials and metamaterials: roadmap,
Optical Materials Express {\bf 15}(7), 1544 (2025).

\bibitem{Theor_Phys_10} Е. М. Лифшиц, Л. П. Питаевский, 
 {\it Физическая кинетика}, Наука, М. (1979). 

\bibitem{RDFC2011} C. Rizza, A. Di Falco, and A. Ciattoni,
%Gain assisted nanocomposite multilayers with near zero permittivity
%modulus at visible frequencies,
Applied Physics Letters {\bf 99}, 221107 (2011).

\bibitem{active_2010}
S. Xiao, V. P. Drachev, A. V. Kildishev, X. Ni, U. K. Chettiar,
H.-K. Yuan, and V. M. Shalaev,
%Loss-free and active optical negative-index metamaterials,
Nature {\bf 466}, 735 (2010).

\bibitem{active_2011}
X. Ni, S. Ishii, M. D. Thoreson, V. M. Shalaev, S. Han, S. Lee, and A. V. Kildishev,
%Loss-compensated and active hyperbolic metamaterials,
Optics Express {\bf 19}, 25242 (2011).

\bibitem{SYG2013} L. Sun, X. Yang,  and J. Gao,
%Loss-compensated broadband epsilon-near-zero metamaterials with gain media
Applied Physics Letters {\bf 103}, 201109 (2013).

\bibitem{K1974} E. A. Kuznetsov,
%The collapse of electromagnetic waves in a plasma,
Zh. Eksp. Teor. Fiz. {\bf 66}, 2037 (1974); [Sov. Phys. JETP {\bf 39}(6), 1003 (1974)].

\bibitem{ZMR1985} V. E. Zakharov, S. L. Musher, and A. M. Rubenchik,
%Hamiltonian approach to the description of non-linear plasma phenomena,
Physic Reports {\bf 129}, 285 (1985).

\bibitem{KRZ1986} E. A. Kuznetsov, A. M. Rubenchik, and V. E. Zakharov,
%Soliton stability in plasmas and hydrodynamics,
Physic Reports {\bf 142}, 103 (1986).

\bibitem{PSSW1991} G. C. Papanicolaou, C. Sulem, P. L. Sulem, and  X. P. Wang,
%Singular solutions of the Zakharov equations for Langmuir turbulence,
Phys. Fluids B {\bf 3}(4), 969 (1991).

\bibitem{HS2009} F. Haas and P. K. Shukla,
% Quantum and classical dynamics of Langmuir wave packets,
Phys. Rev E  {\bf 79}, 066402 (2009).

\end{thebibliography}
\end{document}